\documentclass{icrc29}

\usepackage{graphicx,amssymb,amsmath,times}
\begin{document}

\title{Photodisintegration of cosmic ray nuclei in galaxies and galaxy cluster 
radiation fields}

\author[T. A. Porter \& D. Allard] {T. A. Porter$^a$, D. Allard$^{b,c}$ \\
(a) Department of Physics and Astronomy, Louisiana State University, Baton Rouge, LA70803, USA. \\
(b) Kavli Institute of Cosmological Physics, University of Chicago, 5640 S. Ellis, Chicago, IL60637, USA. 
\\ (c) Department of Astronomy and Astrophysics, University of Chicago, 5640 S. Ellis, Chicago, IL60637, USA. 
}

\presenter{T. A. Porter (tporter@lsu.edu), \
usa-porter-T-abs2-he21-poster}

\maketitle

\begin{abstract}
We present the results of a new calculation for the
photodisintegration of ultra-high energy cosmic ray nuclei
on soft photon targets. We include all the
relevant photodisintegraton processes i.e giant dipole resonance,
quasi-deuteron, baryonic resonances and photofragmentation. In particular,
for the giant dipole process we use a recent calculation of the 
giant dipole cross section. 
We calculate the mean free path for photodisintegration
processes for radiation fields that are likely to exist within galaxies and 
clusters of galaxies.
\end{abstract}

\section{Introduction}
%
%The so-called `bottom-up' scenario of ultra high energy (UHE) cosmic-ray
%acceleration has cosmic-rays being accelerated in the jets emanating from the 
%cores of galaxies.
%An example of a galaxy in which this may be taking place 
%is the giant elliptical M87 (Reimer et al. 2004) in the Virgo cluster.

Galaxies produce diffuse radiation fields by direct stellar
emission, as well as re-processing of starlight where there is sufficient dust.
Furthermore, within galaxy clusters, dust associated with gas 
stripped from galaxies during infall to a cluster core, or dust ejected into 
the intracluster medium by intracluster stars, can result in diffuse emission.
Therefore, in the vicinity of galaxies, and within the intracluster medium, 
there may be additional radiation fields that could increase the total diffuse
radiation field over the best current model for the infra-red (IR)
background in the extragalactic medium (EGM) given 
by Malkan \& Stecker (1998) (MS1998).

The photodisintegration rate of UHE cosmic-ray nuclei depends on the mass 
number 
of the nucleus, $A$, and the soft target photon background.
Photonuclear disintegration cross-sections scale with $A$, making heavier
nuclei more subject to disintegration processes.
Further, changes in the assumed photon background over the EGM background will 
alter the photodisintegration mean free path (MFP).
Specifically, an enhancement of the target background, without altering 
the spectral shape, would correspondingly lower the MFP.
If the spectral shape of any additional target background differs from
EGM background, the importance of the various photonuclear processes 
(giant dipole resonance, 
quasi-deuteron, baryonic resonances, and photo-fragmenation; 
see later) may be altered.

To investigate whether these likely additional diffuse photon backgrounds
can be of importance in UHE cosmic-ray propagation, we use a new calculation 
of photodisintegration of 
UHE cosmic ray nuclei on soft photon targets. 
Together with models for the radiation fields of galaxies and dust
within clusters, we calculate the MFP for 
photodisintegration in the vicinity of galaxies, and the intracluster medium, 
and discuss the possiblity of enhancements to the photodisintegration rate of
UHE nuclei resulting from these.

\section{Interactions of Cosmic Ray Nuclei}
Nuclei are subject to photo-erosion, i.e., they lose nucleons through 
photonuclear interactions mainly with the cosmic microwave background (CMBR), 
IR, and optical backgrounds.  
This involves the following processes: 
i) giant dipole resonance (GDR), with a loss of one or more nucleons, as well 
as $\alpha$-particles 
(this process occurs with photons above a threshold of $\sim 8$~MeV in the 
nucleus rest frame (NFR)), 
ii) the quasi-deuteron (QD) process, where a virtual pion interacts with a 
nucleon pair within 
the nucleus, leading to the ejection of the pair and possibly additional 
protons or neutrons (this occurs with photons of typically $20-30$~MeV in the 
NFR), 
iii) baryonic resonances (BR), where a real pion is produced, ejecting a 
nucleon and possibly interacting further with a nucleon pair, 
eventually leading to the loss, on average, of six nucleons for an iron 
nucleus (this occurs for photons above $\sim 150$~MeV in the NFR), 
and iv) the photo-fragmentation (Frag), occurring at very high energy 
(photons with energies $\sim 1$~GeV in the NFR) and breaking the nucleus into 
many fragments of much lower mass and energy.

For the treatment of the GDR cross sections we apply the revised scheme of 
photonuclear interactions described by Kahn et al. (2005), 
whereas for the higher 
energy processes we use the phenomenological parameterisation obtained 
by Rachen (1996) (see also Allard 2004).  

%\begin{figure*}[ht]
%\centering
%\hfill~\hfill\includegraphics[width=7.5cm]{Backgrounds.eps}
%\hfill\includegraphics[width=7.5cm]{LPM_IR_GC_processes.eps}
%\hfill~\caption{Left (a): Spectral density as a function of the 
%energy for the backgrounds we consider (see text). Right (b): 
%Contribution of the 
%different photonuclear processes to the total MFP as a function of the 
%Lorentz factor of an iron nucleus for the infrared background at the centre of
%our spiral galaxy model.}\label{fig:backgroundsMFP}
%\end{figure*}
\section{Diffuse Photon Fields}
Galaxy clusters typically contain a mixture of galaxy types : ellipticals, 
spirals, and irregulars.
Additionally, dust stripped from galaxies during infall, or ejected from 
intergalactic stars in the cluster, may be present in the 
intracluster medium (Popescu et al. 2000).
All of these may emit diffuse radiation, and we describe the model spectra
we use to represent these different sources of radiation.

For the radiation field in spiral galaxies, we use the 
results of Porter \& Strong (2005).
Their code allows the radiation field to be calculated at all points within a
specified volume in and surrounding a galaxy.
By specifying a distribution of stars and dust, the radiation field can be 
efficiently calculated from the optical to IR.
Absorption and scattering by the assumed dust model is taken into account, 
and a heating code calculates the IR emission by grains undergoing
transient and equilibrium heating in the interstellar medium.
We use the stellar and dust models they assume for the Milky Way, and a
cylindrical geometry (with symmetry about the galactic plane and in azimuth).
We calculate the radiation field for two representative positions : 
the galactic centre (GC), and for a distance 50 kpc from the GC.
The spectrum we obtain for the Milky Way is similar to the spectra 
Popescu \& Tuffs (2002) obtain for late-type spirals in the Virgo cluster.

We use the results of Mazzei et al. (1994) for our 
model of the radiation field in elliptical galaxies.
Their model includes a stellar distribution obtained from an evolution
code, and dust emission by two dust components : warm dust heated in 
regions of high radiation intensity (e.g. OB clusters), and cold dust heated by
the general interstellar radiation field. 
We adopt their results for the most evolved model they consider ($T = 15$ GYr, 
their Fig. 4; see also their Table 1).
We take the spectrum from their model to represent the radiation field 
at a distance $\sim 10$ kpc from the elliptical GC.
We obtain spectra for the GC and a distance 50 kpc from the GC by 
multiplying by factors 100 (GC), and $1/25$ (50 kpc) respectively.
%We take their model to be the spectrum averaged over the 
%central $\sim 10$ kpc of the galaxy.
%We scale this radiation field by a factor $\sim1/25$ to obtain a 
%representation for the spectrum also for a distance $\sim 50$ kpc from the 
%elliptical GC.
Naturally, the size range for ellipticals is quite large, and we would not 
expect this model to be a completely realistic representation 
across the full size range, but it should be sufficient for our current 
purposes.

\begin{figure*}[hbt]
\centering
\hfill~\hfill\includegraphics[width=6.5cm]{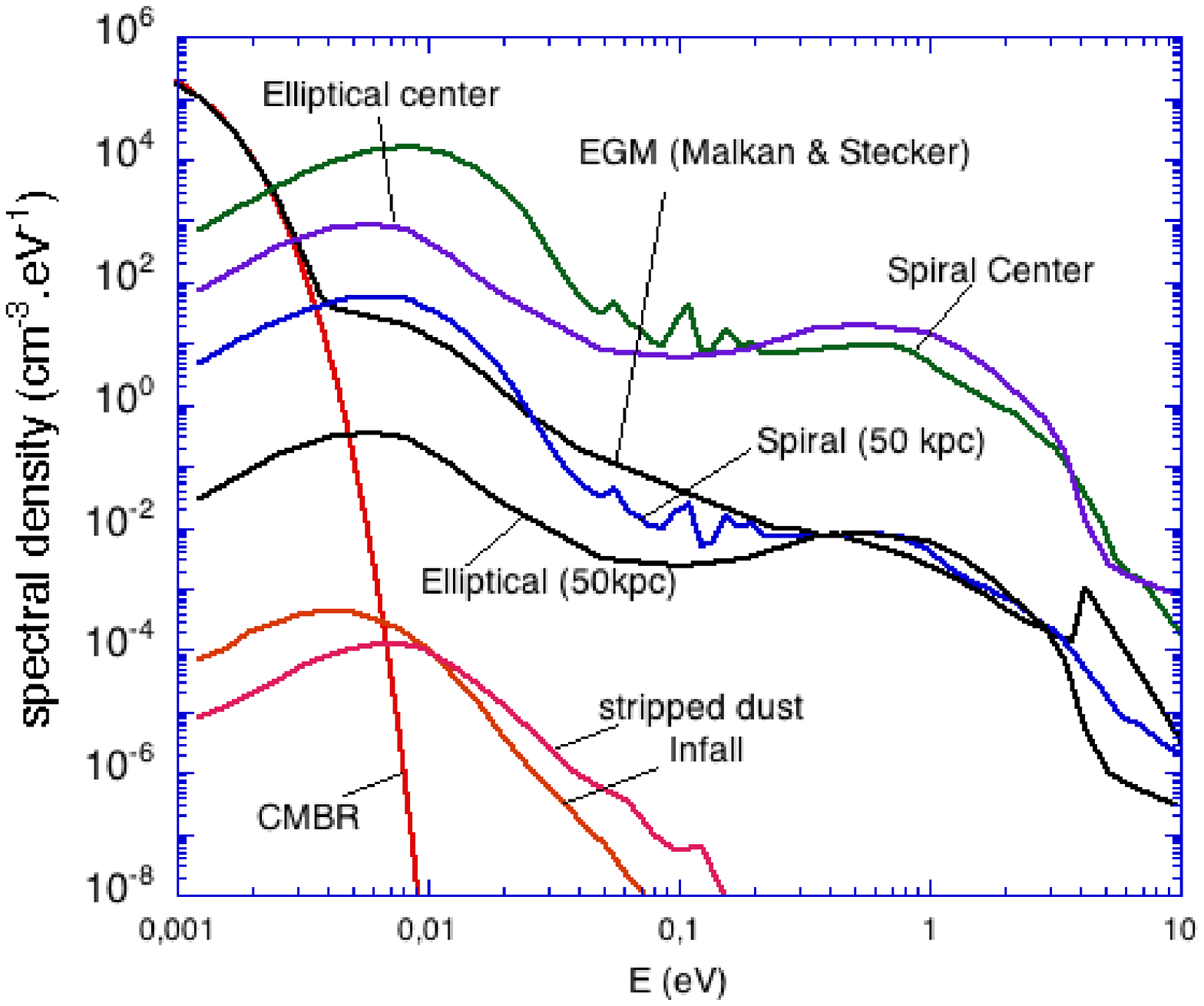}
\hfill\includegraphics[width=6.5cm]{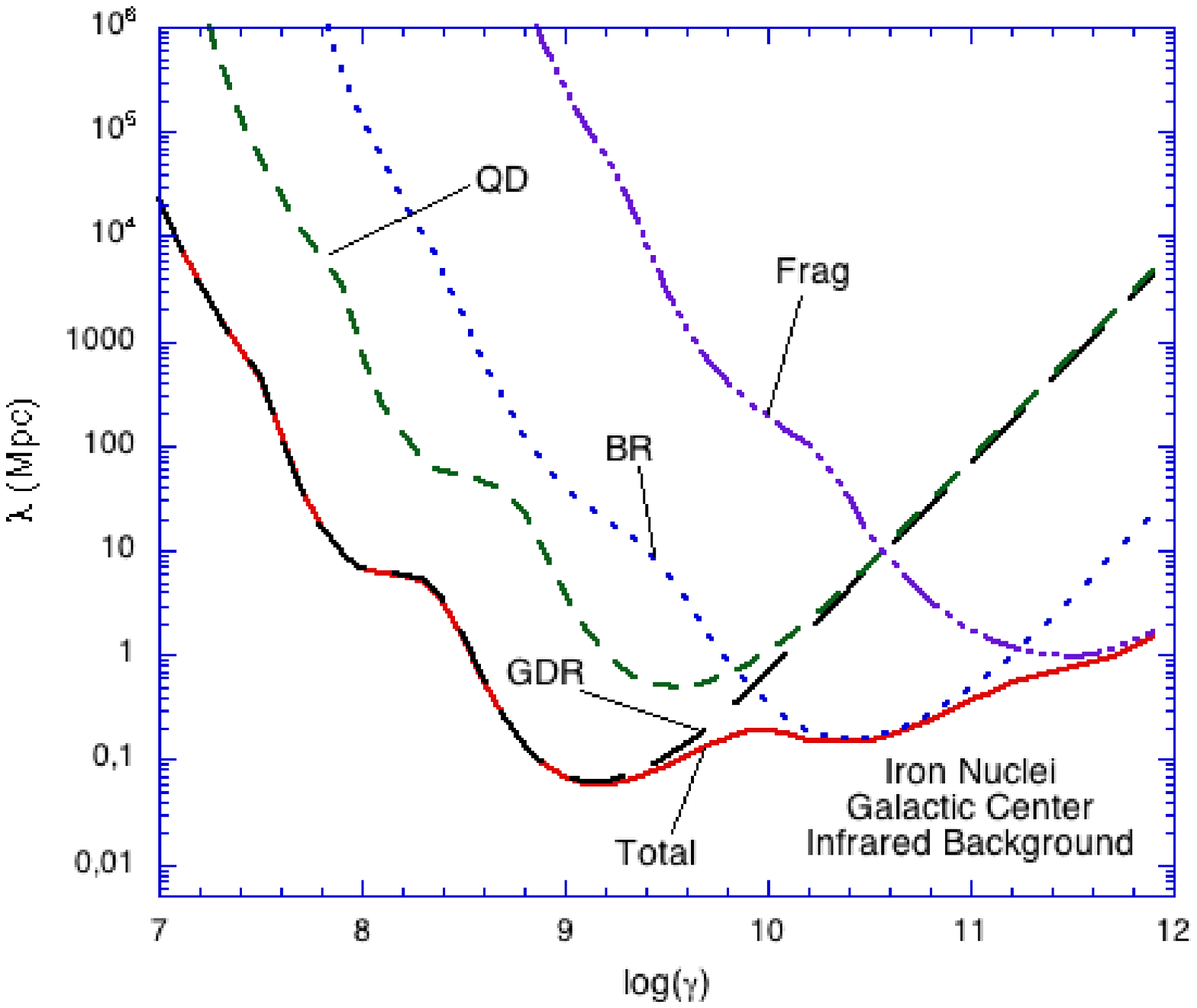}
\hfill~\caption{Left (a): Spectral density as a function of the 
energy for the backgrounds we consider (see text). Right (b): 
Contribution of the 
different photonuclear processes to the total MFP as a function of the 
Lorentz factor of an iron nucleus for the infrared background at the centre of
our spiral galaxy model.}\label{fig:backgroundsMFP}
\end{figure*}

For the intracluster dust spectra, we use the results of Popescu et al. (2000).
In Fig.~\ref{fig:backgroundsMFP}a we show the sectral density of our assumed 
radiation fields 
for spirals, ellipticals, and dust in the intracluster medium. 
Also shown is the MS1998 EGM background.
We can see immediately that the emission by intracluster dust is far too 
low to be significant. 
However, the intensity of the radiation fields for spiral and elliptical 
galaxies is considerable at the GC, and still comparable with the 
EGM background at $\sim 50$ kpc from the GC regions.
%even if 
%the infrared background is in general weaker in ellipticals due to a lower 
%dust density.%can you write something better%

%\begin{figure*}[]
%\centering
%\hfill~\hfill\includegraphics[width=6.5cm]{Backgrounds.eps}
%\hfill\includegraphics[width=6.5cm]{LPM_IR_GC_processes.eps}
%\hfill~\caption{Left (a): Spectral density as a function of the 
%energy for the backgrounds we consider (see text). Right (b): 
%Contribution of the 
%different photonuclear processes to the total MFP as a function of the 
%Lorentz factor of an iron nucleus for the infrared background at the centre of
%our spiral galaxy model.}\label{fig:backgroundsMFP}
%\end{figure*}

%\begin{figure*}[ht]
%\centering
%\hfill~\hfill\includegraphics[width=7.5cm]{LPMGalacticCenter.eps}
%\hfill\includegraphics[width=7.5cm]{LPMMediumComp.eps}
%\hfill~\caption{Left: Contribution of the different backgrounds at the 
%Galactic center to the total MFP as a function of the Lorentz factor of an 
%iron nucleus. Right: Comparison of the MFPs in different media 
%(the contribution of the CMBR and the extragalactic radiation field is added 
%for the MFPs of spirals and ellipticals).}
%\label{fig:backgroundsMFP2}
%\end{figure*}

\section{Photodisintegration Mean Free Paths}
Figure \ref{fig:backgroundsMFP}b shows our calculated MFP in the GC 
IR background for the spiral galaxy model, for the different 
photodisintegration processes. 
We can see that the GDR and BR processes give the main contribution to the 
total MFP for Lorentz factors up to $\sim 4 \times 10^{11}$. 
Above this, fragmentation becomes the leading contributor to the MFP.
From the figure, the effect of the different energy thresholds for
the various photodisintegration processes can clearly be seen.
%The minimum value for the BR is shifted to higher energy due to 
%the higher energy threshold of this process. 
The total MFPs therefore exhibit a complex shape, with several features
due to the contributions of the different processes.
This is true whatever background spectrum is considered.
%The total MFPs (this is true whatever the background we consider) thus 
%exhibit a complex shape with several features due to the contributions of 
%the different processes. 

The background radiation in our galaxy models 
is the sum of the contributions by the
CMBR, IR, and optical backgrounds.
%In addition, the radiation background at the galactic center is the sum of 
%the CMBR, the infrared and the optical backgrounds. 
In Fig.~\ref{fig:backgroundsMFP2}a we display the contribution of 
these backgrounds (considering all the processes mentioned above) to the 
total MFP in the spiral galaxy model GC. 
We clearly see that the different 
radiation fields have a successively dominant contribution according to their 
spectral density peak energy range. 
In the case of the GC spectrum, the MFPs are extremely small in the 
whole Lorentz factor range. 

The magnitude of the radiation field for either of our galaxy models will 
diminish with distance from their respective GCs.
At some point, the intensity will become similar to that of the EGM background.
From Fig.~\ref{fig:backgroundsMFP}a, this is $\sim 50$ kpc for both models. 
Correspondingly, we would expect the MFPs calculated for either galaxy model
to rise with increasing distance from the centre region.
In Fig.~\ref{fig:backgroundsMFP2}b we show the MFPs calculated for our
galaxy models for the GC region and 50 kpc from the GC, 
along with the EGM background MFP.
The MFPs are significantly increased at 50 kpc when compared 
to those calculated for the GCs. 
%Note that the EGM background is included in each of the MFPs calculated 
%using the 
%galaxy radiation fields. 

\section{Discussion}
We have shown that the radiation fields within, and 
in the vicinity of, galaxies can cause a decrease in the MFPs for 
photodisintegration processes. 
The intracluster dust, on the other hand, provides virtually no 
modification to the photodisintegration MFPs.

At the centres of galaxies, the radiation field is significantly larger 
than the EGM background and, therefore, the photodisintegration MFP is very 
small.
However, even if the magnetic fields are greatly enhanced in the centre
of galaxies, UHE particles cannot remain confined for a sufficiently long
time to be signficantly affected by photodisintegration processes. 
This remains true even if a whole galaxy comparable to our spiral model
is considered where the IR background is quite high in the disc, especially 
in the spiral arms.
%This effect is less significant 
%in the case of ellipticals since the IR background. 
%One should anyway keep in mind that some ellipticals are known to have a 
%high density of dust in their disk %can you write something better%.  
%However, even if the magnetic fields are enhanced in the center of 
%galaxies the high energy particles cannot remain confined a 
%sufficently long time to be significantly affected by 
%photodesintegration processes. 
%This remains true if one considers a whole galaxy comparable with ours 
%even if the infrared background is very high in the disk 
%especially in the spiral arms. 

\begin{figure*}[t]
\centering
\hfill~\hfill\includegraphics[width=6.5cm]{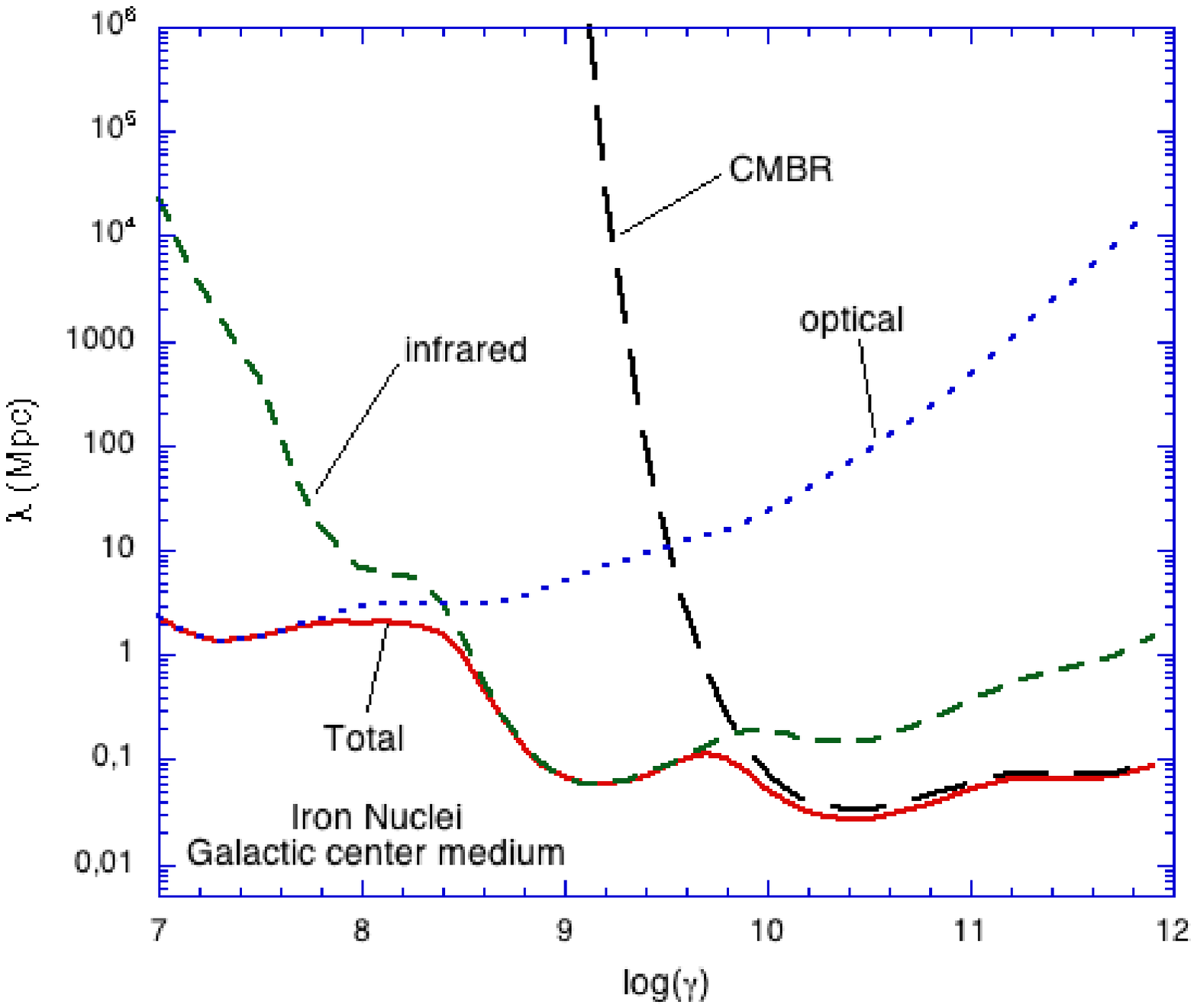}
\hfill\includegraphics[width=6.5cm]{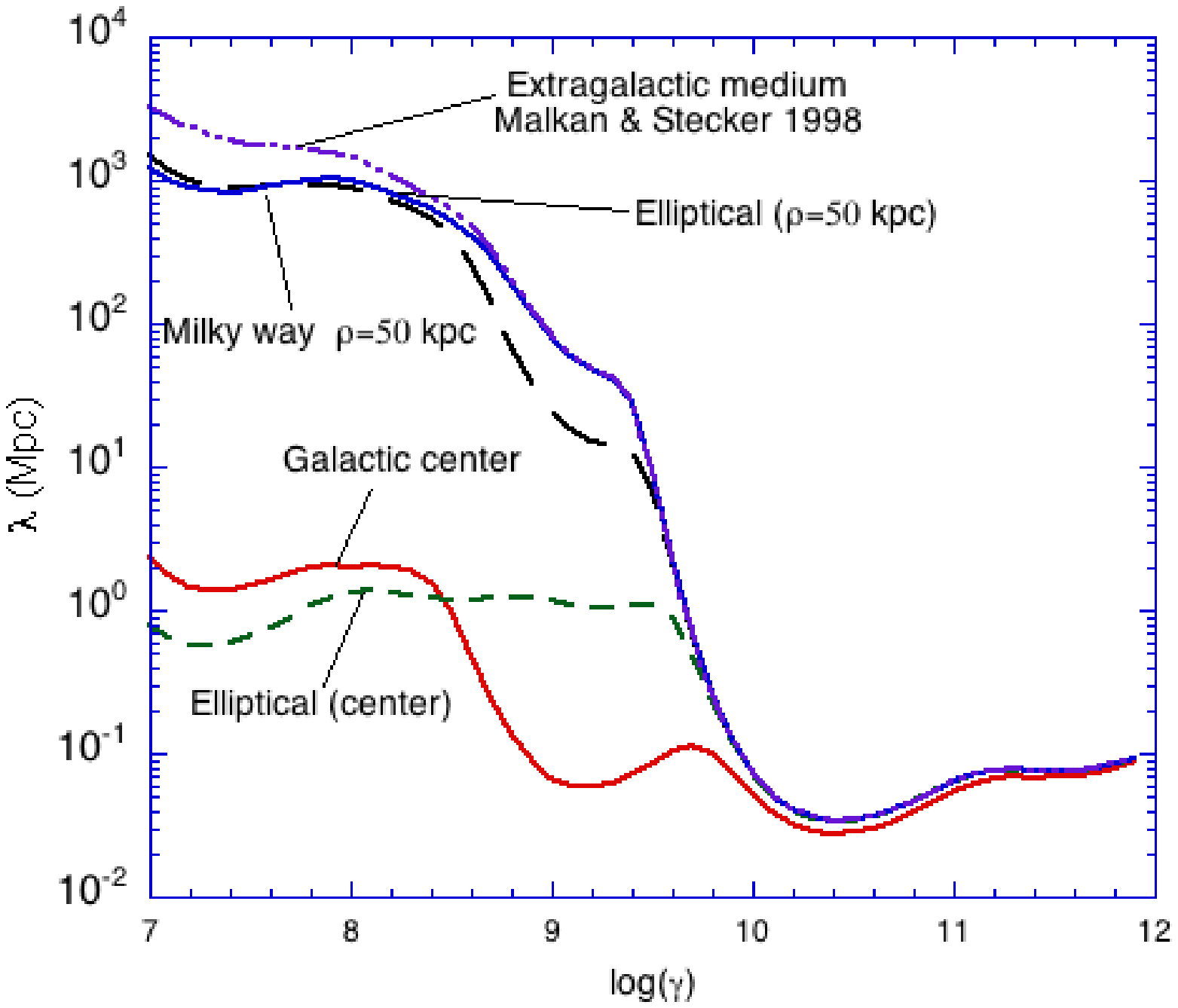}
\hfill~\caption{Left: Contribution of the different backgrounds at the 
Galactic center to the total MFP as a function of the Lorentz factor of an 
iron nucleus. Right: Comparison of the MFPs in different media 
(the contribution of the CMBR and the extragalactic radiation field is added 
for the MFPs of spirals and ellipticals).}
\label{fig:backgroundsMFP2}
\end{figure*}

However, when we consider the region in which the MFP is lower than that for
just the pure EGM background, the situation is more interesting. 
Figure~\ref{fig:backgroundsMFP2}a shows 
a region $\sim 50$ kpc in radius has 
a MFP that is lower than the pure EGM background.
Generally, the size of this region will depend on the luminosity of the galaxy,
its dust content, and its size. 
%Therefore, in the vicinity of galaxies bigger than our models, we could have 
%an even further region of enhanced radiation field, and hence lower 
%photodisintegration MFP.
%For smaller galaxies than our models, the converse is obviously the case.
%However, since in galaxy clusters the typical between galaxy is 
%typically of the order or less than 100 kpc it is even more 
%interesting to study the influence of the extended radiation 
%background on the propagation of high energy nuclei. 
%As we can see in Fig.~\ref{fig:backgroundsMFP2}a, at 50 kpc 
%from the the center the Galactic background is comparable with the 
%extragalactic background on the whole energy spectrum and this is also 
%true for elliptical in the optical range. 
%The corresponding MFPs are displayed on Fig.~\ref{fig:backgroundsMFP2}b and 
%compared with the MFPs in central regions of spirals and eliptical and in 
%the extragalactic medium (EGM). 
%We see that the MFP are significantly lower at 50 kpc from the 
%center of the Galaxy than in the EGM. This effect is less significant 
%in the case of ellipticals since the IR background. 
%One should anyway keep in mind that some ellipticals are known to have a 
%high density of dust in their disk %can you write something better%.  
This seems to indicate that the interaction rate should be higher
in the intracluster medium than in the EGM.
%Dependent on the distance between galaxies in a cluster, cluster morphology, 
%and the physical parameters of galaxies, significant enhancements to the 
%photodisintegration rate could occur within the intracluster medium.

Having shown that the extended radiation fields of galaxies can alter the 
photodisintegration rate above the usual EGM rate, we intend to study this 
further.
We will perform a study, carefully incorporating cluster structure, 
intracluster galaxy distances, and galaxy properties.
This forthcoming study will be of particular interest for UHE cosmic-ray 
propagation, since 
%UHE nuclei could be accelerated in some region of a galaxy
%cluster, e.g., in the jets of active galaxies.
the presence of high radiation fields, combined with the high magnetic 
fields expected in the intracluster medium, could prevent the UHE nuclei
from escaping without significant interactions and energy losses, and 
could also lead to a significant neutrino emission sub-product of the 
neutron emission from photodisintegration.
%
%, dependent on the distance between galaxies in a cluster, 
%cluster morphology, 
%and the physical parameters of galaxies, significant enhancements to the 
%photodisintegration rate could occur within the intracluster medium.
%, we intend to investigate this further.
%
%\section{Discussion}
%We have shown that the extended infrared and optical background lead to an 
%increase of iron interaction rate in the vicinity of spiral and ellipticals 
%galaxies. 
%This seems to indicate that the interaction rate should be significantly 
%higher in the intra-cluster media than in the EGM. 
%However in order to precisely estimate this effect, one has to carrefully 
%take into account the distribution of spirals and ellipticals, their size 
%luminosity and dust content.
%This forthcoming study is of particular interest for UHE cosmic rays since 
%they could be accelerated in some region of galaxy cluster for instance in 
%the jets of active galaxies. 
%The presence of high radiation fields (combined with the high magnetic 
%fields expected in intra-cluster media) could prevent UHE nuclei to escape 
%without interactions and energy loses and also lead to a significant neutrino 
%emission sub-product of neutron emission from photodisintegration.
%


\begin{thebibliography}{99}

\bibitem{Allard04} Allard, D.,  2004, PhD thesis, University Paris VI

%\bibitem{boldt2000} Boldt, E. \& Loewenstein, M., MNRAS 316L, 29 (2000).

%\bibitem{jerjen2004} Jerjen, H., Binggeli, B., \& Barazza, F. D., 
%ApJ 127, 771 (2004).

\bibitem{Khan+05} Khan, E., et al., 2005, Astropart. Phys., 23, 191

\bibitem{malkan1998} Malkan, M. A. \& Stecker, F. W., ApJ 496, 13 (1998).

%\bibitem{malkan2001} Malkan, M. A. \& Stecker, F. W., ApJ 555, 641 (2001).

\bibitem{mazzei1994} Mazzei, P., De Zotti, G., \& Xu, C., ApJ 422, 81 (1994).

\bibitem{popescu2000} Popescu, C. C., et al., A\&A 354, 480 (2000).

\bibitem{popsecu2002} Popescu, C. C. \& Tuffs, R. J., MNRAS 335, 41L (2002).

\bibitem{porter2005} Porter, T. A. \& Strong, A. W., OG 2.1, \emph{this conference} (2005).

\bibitem{Rachen96} Rachen, J. P., \emph{Interaction processes and statistical properties of the propagation of cosmic-rays in photon backgrounds}, PhD thesis of the Bonn University, 1996

%\bibitem{neilsen2000} Neilsen, E. H. Jr. \& Tsvetanov, Z. I., ApJ 536, 255 
(2000).

%\bibitem{reimer2004} Reimer, A., et al., A\&A 419, 89 (2004).

%\bibitem{stecker1999} Stecker, F. W. \& Salamon, M. H., ApJ 512, 521 (1999).

\end{thebibliography}
\end{document}